\documentclass[twocolumn]{aastex63}
\usepackage[fleqn]{amsmath}
\usepackage{float}
\renewcommand{\ng}{NGC\,4258 }
\newcommand{\ngs}{NGC\,4258}
\newcommand{\h}{H$_0$ }
\newcommand{\hs}{H$_0$}
\newcommand{\hst}{{\it HST} }
\newcommand{\hsts}{{\it HST}}
\newcommand{\jwst}{{\it JWST} }
\newcommand{\jwsts}{{\it JWST}}
\newcommand{\hstw}{{\it F350LP} }
\newcommand{\hstv}{{\it F555W} }
\newcommand{\hsti}{{\it F814W} }
\newcommand{\hsth}{{\it F160W} }

\newcommand{\hstws}{{\it F350LP}}
\newcommand{\hstvs}{{\it F555W}}
\newcommand{\hstis}{{\it F814W}}
\newcommand{\hsths}{{\it F160W}}
\newcommand{\hstbs}{{\it F435W}}
\newcommand{\lcdm}{$\Lambda$CDM }
\newcommand{\wesen}{$W_{VI}^{HST}$ }
\newcommand{\wesens}{$W_{VI}^{HST}$}

\makeatletter
\newcommand*{\rom}[1]{\expandafter\@slowromancap\romannumeral #1@}
\makeatother
\shorttitle{Cepheid PLRs in \ngs}
\shortauthors{Yuan et al.}

\begin{document}
\title{Absolute Calibration of Cepheid Period--Luminosity Relations in \ngs}

\author[0000-0001-9420-6525]{Wenlong Yuan}
\affiliation{Department of Physics \& Astronomy, Johns Hopkins University, Baltimore, MD 21218, USA}

\author[0000-0002-1775-4859]{Lucas M.~Macri}
\affiliation{George P.\ and Cynthia W.\ Mitchell Institute for Fundamental Physics and Astronomy,\\ Department of Physics and Astronomy, Texas A\&M University, College Station, TX 77843, USA}

\author[0000-0002-6124-1196]{Adam G.~Riess}
\affiliation{Department of Physics \& Astronomy, Johns Hopkins University, Baltimore, MD 21218, USA}
\affiliation{Space Telescope Science Institute, 3700 San Martin Drive, Baltimore, MD 21218, USA}

\author[0000-0001-5955-2502]{Thomas G.~Brink}
\affiliation{Department of Astronomy, University of California, Berkeley, CA 94720-3411, USA}

\author{Stefano Casertano}
\affiliation{Space Telescope Science Institute, 3700 San Martin Drive, Baltimore, MD 21218, USA}

\author[0000-0003-3460-0103]{Alexei V.~Filippenko}
\affiliation{Department of Astronomy, University of California, Berkeley, CA 94720-3411, USA}
\affiliation{Miller Institute for Basic Research in Science, University of California, Berkeley, CA 94720, USA}

\author[0000-0002-4312-7015]{Samantha L.~Hoffmann}
\affiliation{Space Telescope Science Institute, 3700 San Martin Drive, Baltimore, MD 21218, USA}

\author[0000-0001-6169-8586]{Caroline D.~Huang}
\affiliation{Harvard-Smithsonian Center for Astrophysics, 60 Garden Street, MS-66, Cambridge, MA 02138, USA}

\author[0000-0002-4934-5849]{Dan Scolnic}
\affiliation{Department of Physics, Duke University, Durham, NC 27708, USA}

\begin{abstract}
\ng is one of the most important anchors for calibrating the Cepheid period--luminosity relations (PLRs) owing to its accurate distance measured from water maser motions. We expand on previous efforts and carry out a new Cepheid search in this system using the {\it Hubble Space Telescope} (\hsts). We discover and measure a sample of 669 Cepheids in four new and two archival \ng fields, doubling the number of known Cepheids in this galaxy and obtaining an absolute calibration of their optical PLRs. We determine a Wesenheit (\wesens) PLR of $-2.574(\pm0.034) -3.294(\pm0.042) \log P$, consistent with an independent Large Magellanic Cloud (LMC) calibration at the level of $0.032\pm0.044$~mag in its zeropoint, after accounting for a metallicity dependence of $-0.20\pm0.05$~mag\,dex$^{-1}$ \citep{2016ApJ...826...56R}. Our determination of the PLR slope also agrees with the LMC-based value within their uncertainties. We attempt to characterize the metallicity effect of Cepheid PLRs using only the \ng sample, but a relatively narrow span of abundances limits our sensitivity and yields a \wesen zero-point dependence of $-0.07 \pm 0.21$ mag\,dex$^{-1}$. The Cepheid measurements presented in this study have been used as part of the data to derive the Hubble constant in a companion paper by the SH0ES team.

\ \par
\end{abstract}

\section{Introduction}

The Cepheid period--luminosity relations \citep[PLRs;][]{1912HarCi.173....1L} have played a critical role in determining the extragalactic distance scale and the expansion rate of the Universe. By searching for Cepheid variable stars in M33 and measuring their periods and magnitudes, \citet{1926ApJ....63..236H} confirmed the extragalactic nature of ``spiral nebulae.'' Using Cepheid PLRs and other techniques (brightest stars and mean nebular luminosity), \citet{1929PNAS...15..168H} was able to estimate the current expansion rate of the Universe, now known as the Hubble constant (\hs). It is one of the few directly observable cosmological parameters that constrain the age, size, and evolution of our Universe. Since then, many dedicated studies \citep[e.g.,][]{2001ApJ...553...47F} leveraged the robustness of Cepheid PLRs to calibrate the luminosity of various secondary distance indicators such as Type Ia supernovae (SNe Ia), the Tully-Fisher relation, the surface brightness fluctuation method, and the fundamental plane for elliptical galaxies, to determine \h using their fluxes and redshifts in the Hubble flow. SNe Ia are unrivaled among secondary distance indicators in their accuracy and precision for studies of \h and dark energy \citep{2018ApJ...859..101S}.

In the past decades, observational cosmologists have sought ever-greater precision in \h for a better understanding of our Universe. As key steps in the measurement of \hs, the calibration and application of Cepheid PLRs are now pushed to a precision that requires propagation of uncertainties from subtle factors besides the random error in the astrophysically-limited sample size of Cepheid--SN~Ia calibrators, such as differences in reddening laws, metallicity scales, possible PLR nonlinearities, and instrument calibrations. The SH0ES (Supernovae, \hs, for the Equation of State of dark energy) team provides the ``state-of-the-art'' measurements of the Cepheid-based distance ladder while observing the greatest number of SN~Ia hosts leading to the most precise direct measurement of \h \citep{2021arXiv211204510R} through careful designs of observations.  This includes the use of the same {\it Hubble Space Telescope} (\hsts) instruments and filters to measure Cepheids in all galaxies to nullify cross-instrumental errors, the pre-selection of low-dust, ideal SN~Ia calibrators, and the choice of near-infrared (NIR) wavelengths to reduce uncertainties from knowledge of reddening laws. After more than a decade of work \citep{2009ApJ...699..539R,2011ApJ...730..119R,2016ApJ...826...56R,2019ApJ...876...85R,2021arXiv211204510R}, the SH0ES team is approaching 1\% total uncertainty in \h and has discovered a discrepancy between the direct measurement of \h and its inferred value from the combination of {\it Planck} cosmic microwave background observations and the \lcdm cosmological model that is $5\sigma$ in significance \citep{2021arXiv211204510R}. Extensive checks of systematic errors in both the direct measurement and the {\it Planck} observations do not relieve this tension \citep[][and references therein]{2021arXiv211204510R,2021CQGra..38o3001D}, raising the possibility of ``new physics'' beyond the standard model.

In the current stage, the Cepheid PLR zeropoints are anchored to three systems: the Large Magellanic Cloud (LMC), the Milky Way, and \ngs. The distances to these three anchors were all measured geometrically, with the LMC distance obtained using detached eclipsing binaries, the \ng distance derived through monitoring of its water-maser motions, and the Milky Way Cepheid distances measured via  parallaxes from the {\it Gaia} mission as well as \hst via the spatial-scanning technique. The past three years have witnessed major improvements in all three of these distance measurements. \citet{2019Natur.567..200P} measured 20 eclipsing binary systems and achieved a 1\% distance measurement to the LMC. \citet{2019ApJ...886L..27R} improved the water-maser modeling and reduced the \ng distance uncertainty from 2.6\% to 1.5\%. For the Milky Way Cepheids, the {\it Gaia} EDR3 release delivered a factor of 2 higher precision than DR2. Concurrently with these improvements of anchor distances, the SH0ES team \citep{2019ApJ...876...85R,2021ApJ...908L...6R} acquired \hst observations of 70 and 75 Cepheids in the LMC and Milky Way, respectively, to calibrate the Cepheid PLRs with the same observational setup as used for the Cepheid--SN~Ia calibrator observations, nullifying the cross-instrumental errors in the \h measurement. For \ngs, dedicated \hst observations and analyses were performed by \citet{2006ApJ...652.1133M}, \citet{2016ApJ...830...10H}, and \citet{2016ApJ...826...56R}.

Given the critical role of \ng in the Cepheid PLR calibration and thus the \h measurement, we expanded the Cepheid observations in this galaxy to enlarge the Cepheid sample, especially long-period ($P>10$\,d) Cepheids, which are more commonly used for the Cepheid--SN~Ia calibration. In this study, we observe four new fields in \ng using \hst and perform a uniform Cepheid search and analysis in them, as well as perform a blind search and measurement of Cepheids from two archival fields \citep{2006ApJ...652.1133M}. The rest of this paper is organized as follows. In \S2, we describe the observations of various \ng fields and detail our reduction and photometry procedures. \S3 presents the Cepheid identification and characterization procedures. We show our results in \S4 and summarize the paper in \S5.

\section{Observations and Data Reduction}

\subsection{Observations}

\ng is a relatively nearby spiral galaxy at a distance of 7.58\,Mpc \citep{2019ApJ...886L..27R}. Because of its relatively large apparent size, measuring a substantial sample of long-period Cepheids in this system using \hst (which has imagers with limited fields of view of $\sim 2'$--$3'$ on a side) is less efficient than in more-distant large spirals. Nevertheless, given the critical role of \ng in the absolute calibration of the extragalactic distance scale, time-series \hst observations toward selected fields of this galaxy are desired in order to calibrate Cepheid PLRs in the same filter system as for those more distant Cepheid--SN~Ia calibration galaxies. In 2003--2004, two \ng fields were observed with 12 epochs using \hst ACS (GO-9810; PI Greenhill). Random-phased WFC3 observations, including NIR imaging, were also acquired in programs GO-11570 and 15640 (PI Riess) to aid the Cepheid PLR calibration of these Cepheids. We recently designed and acquired new \hst time-series observations (GO-16198, PI Riess) in four additional fields to search for and measure more Cepheids in \ng to better constrain the absolute calibration of Cepheid PLRs. In this study, we analyzed both the archival observations and these new data at optical wavelengths only. The NIR data reduction and results are presented by \citet{2021arXiv211204510R}.

\begin{figure}
\epsscale{1.2}
\plotone{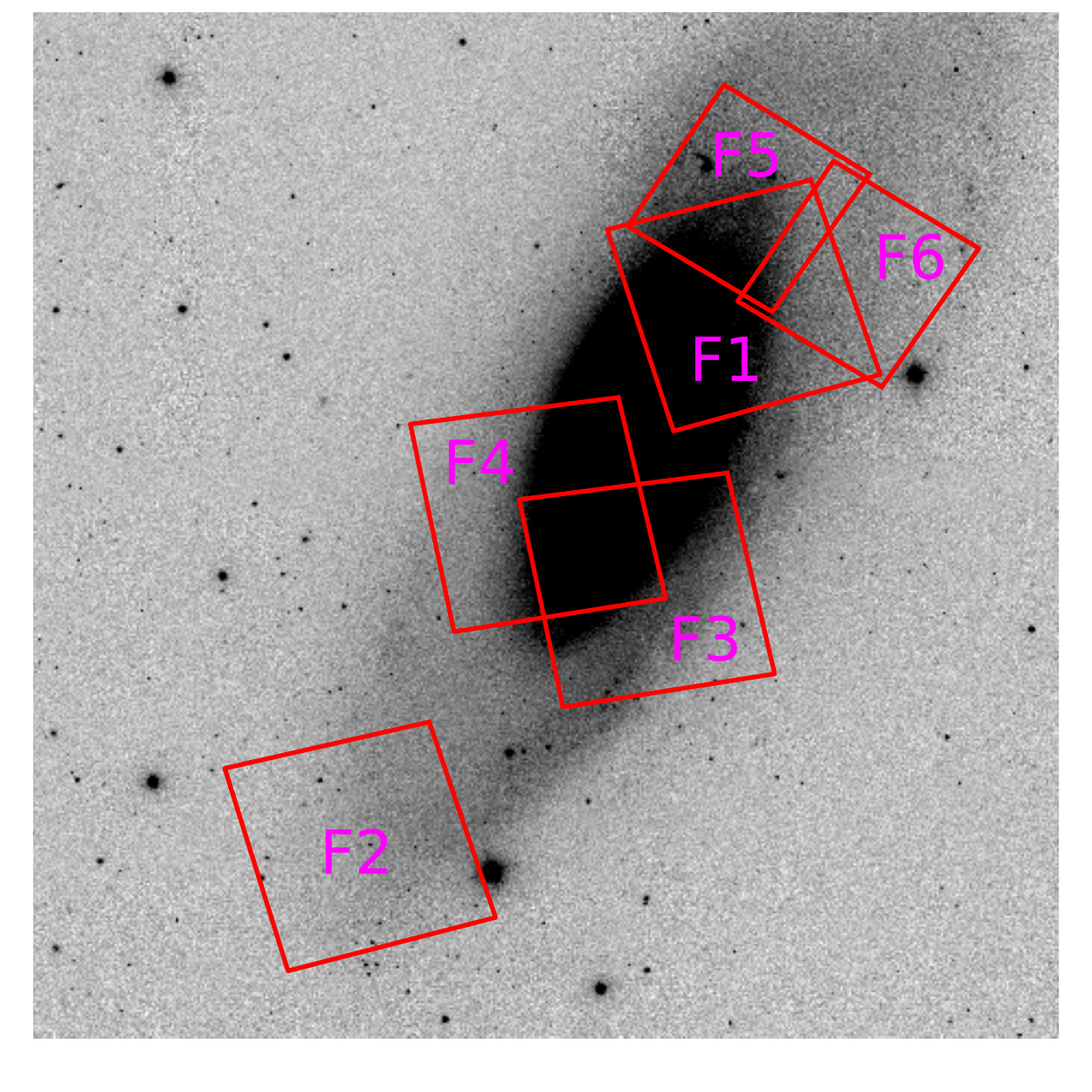}
\caption{Archival (F1 and F2) and new (F3--F6) \hst  observations of \ng using ACS (F1--F4) and WFC3 (F5 and F6) overlaid on an $i$-band Sloan Digital Sky Survey image that is 16.5$\arcmin$ on a side. North is up and east is to the left.}\label{fig_obs}
\end{figure}

\subsubsection{Archival Observations}

The archival observations acquired in 2003--2004 include 12 epochs of ACS imaging data in three bands: \hstbs, \hstvs, and \hsti (\hst equivalent to $B$, $V$, and $I$, respectively). These data were analyzed by \citet{2006ApJ...652.1133M} to search for Cepheids and measure their PLRs. A detailed description of these archival data can be found in \S2.1 of that paper. The observations cover two fields: ``inner'' and ``outer'' as named by the authors. The observation baseline was $\sim 45$ days and two images were obtained in each band of each visit to help remove cosmic rays. In this work, we re-analyzed only the \hstv and \hsti data for their relevancy to the SH0ES calibration of the Cepheid distance scale. The exposure times per epoch were 1600\,s and 800\,s for \hstv and \hstis, respectively. For convenience, we renamed the inner field as ``F1'' and the outer field as ``F2.'' In this re-analysis, we used the latest {\tt Astrodrizzle} software, which was not yet available at the time when the original analysis was performed. Compared with the {\tt Multidrizzle} software that was used prior to its launch, {\tt Astrodrizzle} provides significant improvements in data processing and astrometric information handling \citep{2012drzp.book.....G}.

\subsubsection{New Observations}

From December 2020 to February 2021, we acquired time-series observations of four \ng fields using \hsts. Two partially overlapping fields (F3 and F4) were imaged with ACS, while the other two fields (F5 and F6, also partially overlapping) were simultaneously imaged with WFC3. We adopted the same \hstv and \hsti filters used by the SH0ES team to calibrate Cepheid PLRs. For the WFC3 fields (F5 and F6), we also included \hstw observations for a more efficient Cepheid search and period determination than the traditional \hstv filter. We obtained between 10 and 13 epochs of \hstv or \hstw data that span a baseline of 68 days depending on the field. Unlike the 2003--2004 archival observations, we only obtained one exposure per visit. Consequently, we were not able to remove cosmic rays in individual images (we removed cosmic rays in our combined master images as defined in \S\ref{sec_phot}). The typical exposure times were 640\,s and 785\,s per band for ACS and WFC3 fields, respectively. Because of \hst orientation constraints, there were some frame offsets among different visits. The locations of the first image of each field are shown in Figure~\ref{fig_obs}. We summarize the observation dates, exposure times, and dataset identifiers of both archival and new observations in Table~\ref{tab_obs}.

\begin{deluxetable}{ccccccc}
\tabletypesize{\scriptsize}
\tablecaption{Observation Log\label{tab_obs}}
\tablewidth{0pt}
\tablehead{
\colhead{Field} & \colhead{Epoch} & \colhead{Date} & \colhead{Dataset} & \colhead{Filter} & \colhead{Exposure [s]} & \colhead{Dither}
}
\startdata
 F1 &   1 &   2003-12-05 &    J8R001020 &   \hsti &     800 &   2 \\
 F1 &   1 &   2003-12-05 &    J8R001030 &   \hstv &    1600 &   2 \\
 F1 &   2 &   2003-12-06 &    J8R002020 &   \hsti &     800 &   2 \\
 F1 &   2 &   2003-12-06 &    J8R002030 &   \hstv &    1600 &   2 \\
 F1 &   3 &   2003-12-07 &    J8R003020 &   \hsti &     800 &   2
\enddata
\tablecomments{This table is available in its entirety in machine-readable form.}
\end{deluxetable}

\subsection{Data Reduction}\label{sec_phot}

We retrieved both the archival and new data from the Mikulski Archive for Space Telescopes (MAST) in the {\tt flc} format, which includes flatfield and charge-transfer efficiency (CTE) corrections. We aligned and drizzled \citep{2002PASP..114..144F} the images of a given field using {\tt DrizzlePac 2.2.6}. For each field, we registered all the images to a reference frame (usually the first \hsti image for fields F1--F4 and the first \hstw image for fields F5 and F6) using {\tt tweakreg 1.4.7}. The registered images were then drizzled to distortion-free images in the {\tt drc} format using {\tt Astrodrizzle 3.1.6}. We created drizzled images for each epoch of each band, which were used for time-series photometry in a later step, and deep master images that included data from all epochs for source-detection purposes. We excluded epoch 10 of fields F1 and F2, and epoch 11 of field F2, from the corresponding master images because they were slightly defocused. We note that the defocused epochs were included in the time-series photometry as their zeropoints can be adjusted using field stars to match the other epochs.

We used the {\tt DAOPHOT}/{\tt ALLSTAR}/{\tt ALLFRAME} software suite \citep{1987PASP...99..191S,1994PASP..106..250S} to perform point-spread function (PSF) photometry on the drizzled images. Our photometry procedures were divided into three steps. In the first step, we derived a master source list from those drizzled deep master images. We adopted the standard two-pass source detection and PSF fitting using {\tt DAOPHOT} and {\tt ALLSTAR} as described in the {\tt DAOPHOT} manual. For fields F5 and F6, we used the \hstw master images for source detection owing to their increased depth and wavelength coverage. For the ACS fields (F1--F4) where \hstw is unavailable, we co-added the \hstv and \hsti master images for source detection. Variable sources with mean magnitudes above the detection limit but first-epoch magnitudes below the detection limit should be safely included in our source list, given that our master images consist of the flux of all the epochs. In the second step, we performed time-series PSF photometry on the drizzled images of individual epochs using {\tt ALLFRAME} with the source list obtained in the first step as input. This was done separately for each band to obtain instrumental \hstws, \hstvs, and \hsti magnitudes for each epoch. In the last step, we calibrated the magnitudes into the Vega system by including aperture corrections and the appropriate instrument zeropoints. We determined the aperture corrections by measuring the magnitude differences between our PSF photometry and aperture photometry of $\sim$150 visually selected bright and isolated objects. The Vega magnitude zeropoints were obtained using the STScI ACS Zeropoints Calculator\footnote{https://acszeropoints.stsci.edu} and \citet{2017wfc..rept...14D} for ACS and WFC3, respectively.

\section{Cepheid Selection}

\begin{figure}
\epsscale{1.2}
\plotone{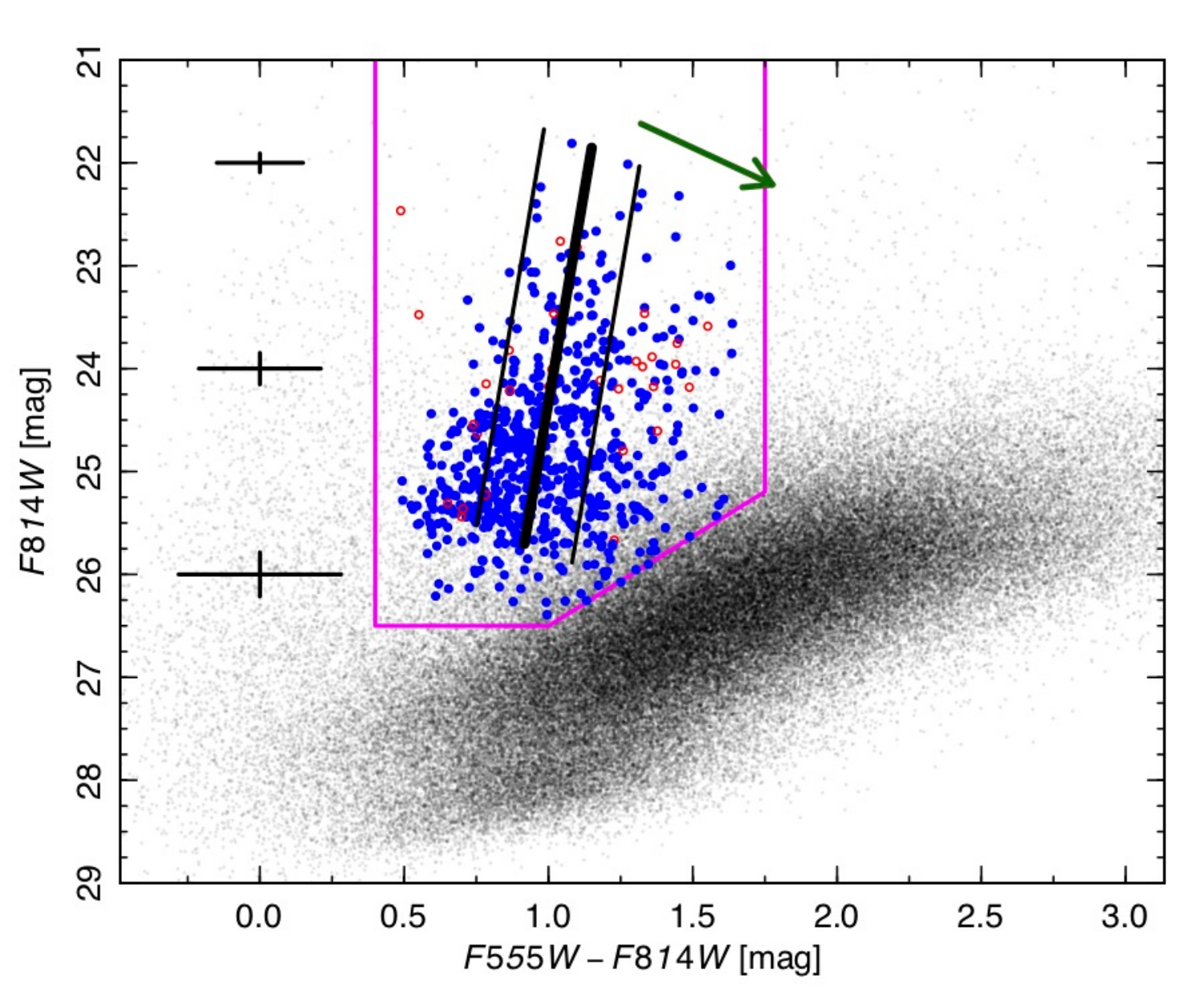}
\caption{Color-magnitude diagram showing Cepheid candidate selection. Magenta lines indicate the initial selection boundary, chosen to encompass the expected Cepheid instability strip while avoiding possible contamination from the dense red giant branch stars. Blue and red points show our final Cepheid sample and rejected outliers, respectively. Representative error bars are shown. The expected mean Cepheid instability strip location and width for $P > 5$\,d, derived from LMC observations \citep{2019ApJ...876...85R}, are shown with thick and thin black lines, respectively. The green arrow indicates the effect of $A_V=1$\,mag. Black dots show only a subset of field stars for visualization purposes.}\label{fig_cmd}
\end{figure}

We detected and measured $\sim 2.8$ million point sources in these six fields using the method described in \S\ref{sec_phot}. To select a clean sample of Cepheid variables, we applied several consecutive cuts based on the light-curve properties, colors, magnitudes, and signal-to-noise ratios (SNRs). The initial cut is based on the light-curve (\hstv for ACS fields and \hstw for WFC3 fields) variability index $L$ \citep{1996PASP..108..851S}. We made use of the {\tt TRIAL} program that was kindly provided by Peter Stetson to extract light curves of sources with $L\!>\!0.5$. The second cut is based on the color--magnitude diagram (CMD) to select stars in a CMD region where the Cepheid instability strip is expected. Our selection boundary is shown with magenta lines in Figure~\ref{fig_cmd}. The lower part of the boundary was designed to avoid the dense red giant branch stars, which would contaminate the selected sample with too many non-Cepheid stars. 

Next, we fit Cepheid templates to the light curves of the surviving objects and made additional cuts. We adopted the \citet{2009AJ....137.4697Y} templates to measure their periods, amplitudes, and mean magnitudes. The details of the template-fitting method are presented in \S3.2 of \citet{2020ApJ...902...26Y}. Then we applied the amplitude-based cuts
\[
\left\{
     \begin{array}{ll}
     A_{\hstv} {\ \textrm{or}\ } A_{\hstw} > 0.13 \: \mathrm{mag}\\
     A_{\hsti} > 0.1 \: \mathrm{mag}\\
     1 < ({A_{\hstv} {\ \textrm{or}\ } A_{\hstw}}) / {A_{\hsti}} < 2
     \end{array}
\right.
\]
and cuts based on the dispersion of light curves about the best-fit models,
\[
\left\{
     \begin{array}{ll}
     \mathrm{F1, F2:}\: \sigma_{\hstv} < 0.15,\: \sigma_{\hsti} < 0.15 \: \mathrm{mag}\\
     \mathrm{F3, F4:}\: \sigma_{\hstv} < 0.18,\: \sigma_{\hsti} < 0.15 \: \mathrm{mag}\\
     \mathrm{F5, F6:}\: \sigma_{\hstw} < 0.15\: \mathrm{mag} .
     \end{array}
\right.
\]
Considering the limited sampling of the \ng light curves, we adopted conservative amplitude-based cuts compared to those well-measured Cepheid amplitude properties as presented by \citet{2009A&A...504..959K}. We note that the Cepheid amplitudes in \hstw and \hstv are close enough \citep{2016ApJ...830...10H} for the above selection purposes. For the cuts based on the light-curve fits, we relaxed the $\sigma_{\hstv}$ cut in fields F3 and F4 owing to their lower-SNR observations compared with other fields and bands. We then removed possible Type \rom{2} Cepheids whose Wesenheit index \wesen are fainter than 28.1 $-$ 3.31$\log P$ or $\sim1.3$\,mag fainter than the classical Cepheid PLR, where \wesen is defined as $\hsti - 1.3(\hstv - \hsti)$ and the Wesenheit color term of 1.3 was derived by \citet{2019ApJ...876...85R} using a reddening law of $R_V=3.1$ from \citet{1989ApJ...345..245C}. We also removed objects with periods shorter than 5 days to avoid incompleteness. We note that this period cut also removes the sample contamination from first-overtone Cepheids. The combination of the above cuts yielded a sample of 699 candidate Cepheids. 

\begin{figure}
\epsscale{1.2}
\plotone{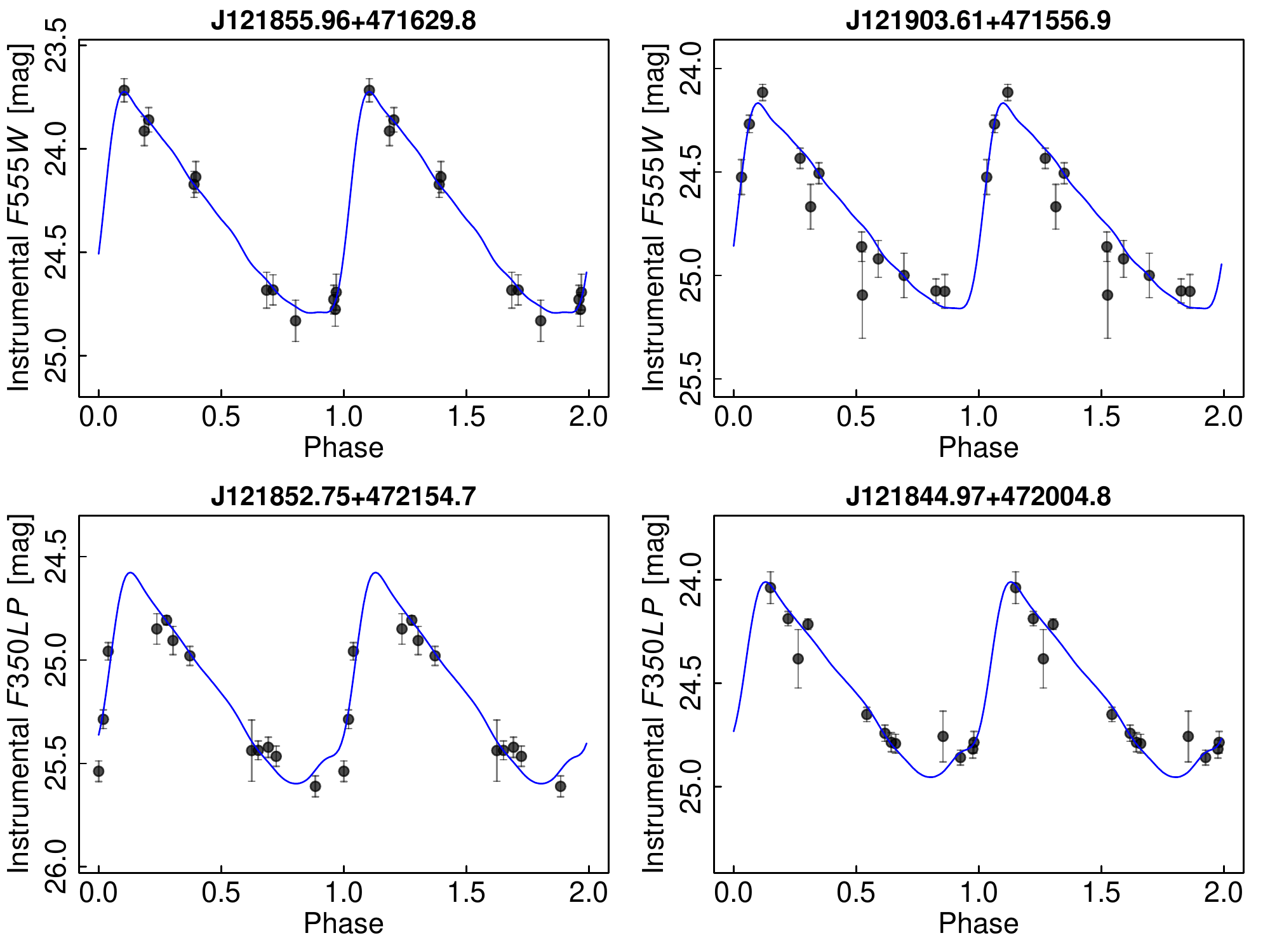}
\caption{Example Cepheid light curves in the new \ng fields (F3--F6). The blue curves show the best-fit \citet{2009AJ....137.4697Y} templates. We plot two phase cycles for visualization purposes.}\label{fig_lcnew}
\end{figure}

\begin{figure}
\epsscale{1.2}
\plotone{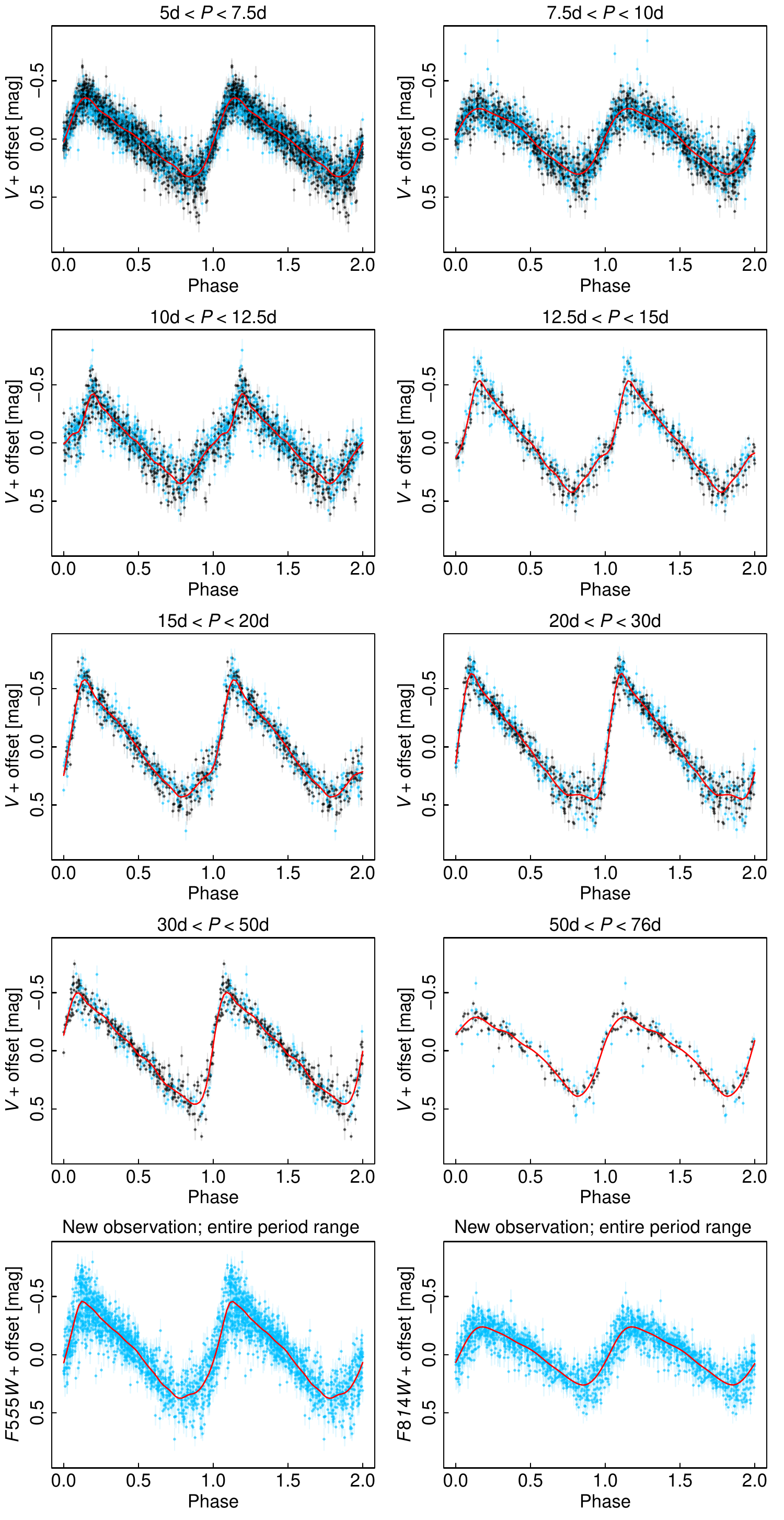}
\caption{Composite light curves of the \ng Cepheids in eight period bins using \hstw or \hstv data (top four rows) and the newly-observed (F3--F6) Cepheids of the entire period range in the \hstv (bottom left) and \hsti (bottom right) bands. Black and blue points indicate archival and new measurements, respectively. Points with uncertainties greater than 0.1\,mag were excluded. Red curves show cubic-spline fits. We plot two phase cycles for visualization purposes. A common phase from the \citet{2009AJ....137.4697Y} templates was imposed.}\label{fig_lc}
\end{figure}
Finally, we cleaned our sample by performing an iterative 3$\sigma$ clipping of their PLRs. We fit the PLRs to the Wesenheit index, \hstvs, and \hsti measurements with fixed slopes of $-3.31$, $-2.76$, and $-2.96$ (respectively) to reject outliers, where the PLR slopes were derived by \citet{2019ApJ...876...85R}. For the \hstv and \hsti bands, we only rejected outliers in the brighter side, which are possibly caused by blending, as the fainter objects are mostly due to extinction and should not be excluded. We rejected 4\% of the objects as outliers in this clipping and derived a final sample of 669 Cepheid candidates. If we only reject Wesenheit outliers, then 13 more objects would survive, shifting the Wesenheit index, \hstvs, and \hsti PLR zeropoints by 0.009, 0.018, and 0.016\,mag, respectively.

We derived background corrections for these objects using the artificial star test method outlined by \citet{2016ApJ...830...10H} and adjusted their final photometry accordingly. By injecting artificial stars near Cepheids and measuring their magnitudes in the same manner as for real-star photometry, we were able to derive and correct any photometry bias due to background variations. The locations and measured properties of our final Cepheid sample are presented in Table~\ref{tab_cep}. Example light curves in new fields (F3--F6) are shown in Figure~\ref{fig_lcnew}, while composite light curves of all fields are shown in Figure~\ref{fig_lc}.

\begin{deluxetable*}{@{\extracolsep{4pt}}lcccccccc@{}}
\tabletypesize{\scriptsize}
\tablecaption{Cepheid Properties\label{tab_cep}}
\tablewidth{0pt}
\tablehead{
\colhead{ID} & \colhead{$P$} & \colhead{R.A.} & \colhead{Decl.} & \multicolumn{3}{c}{Magnitudes$^a$ [mag]} & \multicolumn{2}{c}{Amplitudes$^b$ [mag]} \\ \cline{5-7} \cline{8-9}
& (days) & (J2000.0) & (J2000.0) & \hstv & \hsti & \hstw & $A_V$ & $A_I$
}
\startdata
 J121856.37+471611.8 &     5.010 &   184.734879 &    47.269939 &    26.686(302) &    25.155(355) &    \nodata &     0.298 &     0.214 \\
 J121846.12+472037.4 &     5.021 &   184.692184 &    47.343731 &    26.441(126) &    25.606(166) &    26.354(151) &     0.311 &     0.198 \\
 J121849.57+472117.6 &     5.035 &   184.706543 &    47.354893 &    27.387(190) &    26.392(321) &    27.117(185) &     0.271 &     0.194 \\
 J121904.04+471536.1 &     5.036 &   184.766846 &    47.260014 &    26.642(317) &    25.552(345) &    \nodata &     0.275 &     0.150 \\
 J121854.29+471921.1 &     5.054 &   184.726227 &    47.322521 &    26.578(315) &    25.694(474) &    \nodata &     0.333 &     0.237 \\
 J121848.01+472138.4 &     5.086 &   184.700058 &    47.360661 &    26.129(134) &    25.362(186) &    25.928(134) &     0.418 &     0.247 \\
 J121833.83+472003.3 &     5.086 &   184.640945 &    47.334251 &    25.863(141) &    25.095(199) &    25.968( 93) &     0.277 &     0.198 \\
 J121835.63+472110.9 &     5.087 &   184.648453 &    47.353016 &    26.130(137) &    25.265(213) &    25.966(107) &     0.316 &     0.226 \\
 J121856.58+471617.4 &     5.090 &   184.735764 &    47.271507 &    26.183(359) &    24.940(451) &    \nodata &     0.329 &     0.166 \\
 J121851.55+472108.1 &     5.093 &   184.714783 &    47.352257 &    26.369(242) &    25.634(381) &    \nodata &     0.309 &     0.228
\enddata
\tablecomments{$^a$Fully calibrated Vega magnitudes, including background corrections and their uncertainties. Magnitude uncertainties are shown in parentheses and expressed in units of $10^{-3}$ mag.
$^bA_V$: \hstv or \hstw amplitude. $A_I$: \hsti amplitude.
This table is available in its entirety in machine-readable form.}
\end{deluxetable*}

\section{Results}

\ng is one of the most important distance anchors for Cepheid PLR calibration. The Keplerian motions of water masers around the central supermassive black hole in \ng monitored by Very Long Baseline Interferometry (VLBI) allow a solid geometrical distance measurement of this system with percent-level uncertainty, laying the foundation for absolute calibrations of the Cepheid PLRs in \ngs. In terms of the Cepheid--SN~Ia distance ladder, \ng shows additional advantages. It shares a similar mean crowding level compared to those SN~Ia hosts that the SH0ES team measured, and it exhibits a mean metallicity closer to the SN~Ia hosts than the LMC. Such similarities help reduce possible systematics from crowding corrections and metallicity-dependence corrections.

Multiple works have calibrated the Cepheid PLRs in \ng in various bands. For example, \citet{2001ApJ...553..562N} observed \ng using the WFP2 camera onboard \hst and obtained optical PLRs in $VI$ bands. \citet{2006ApJ...652.1133M} derived optical PLRs in $BVI$ bands using the later-installed ACS camera on \hsts. \citet{2015AJ....149..183H} observed \ng with ground-based facilities and obtained PLRs in the Sloan Digital Sky Survey $gri$ bands. \citet{2015MNRAS.450.3597F} observed \ng using the Large Binocular Telescope and derived PLRs in ground-based $BVI$ bands. \citet{2015MNRAS.450.3597F} and \citet{2016ApJ...826...56R} both reported NIR Cepheid PLRs in the \hst \hsth (equivalent to the ground $H$) band.

In this work, we adopted the latest geometric distance to this system \citep{2019ApJ...886L..27R} and determined the absolute Cepheid PLR zeropoints in optical \hst bands. We compared our zeropoints to those set by LMC Cepheids \citep{2019ApJ...876...85R} and found good agreement to the level of $0.032\pm0.044$\,mag if we assume a metallicity dependence of $-0.20\pm0.05$\,mag\,dex$^{-1}$ derived by \citet{2016ApJ...826...56R}. We also tentatively studied the PLR metallicity effect using the \ng Cepheids. A comprehensive analysis of the metallicity dependence using multiple datasets is beyond the scope of this work; however, we refer interested readers to \citet{2021ApJ...913...38B}, \citet{2021arXiv211204510R}, \citet{2021MNRAS.508.4047R}, and \citet{2022arXiv220101126R} for this subject. A detailed comparison of Cepheid photometry between this work and the many past publications using different cameras and filter systems is also beyond our scope.  However, we note that many past works have not fully incorporated individual Cepheid crowding corrections (e.g., \citealt{2006ApJ...652.1133M}, \citealt{2016ApJ...830...10H}) as we have here, which will make the Cepheids fainter. The Cepheid properties presented in this work were used by \citet{2021arXiv211204510R} to develop the distance ladder and measure \hs. \citet{2021arXiv211204510R} also published the \hsths-band results of these \ng Cepheids.

\begin{figure}
\epsscale{1.2}
\plotone{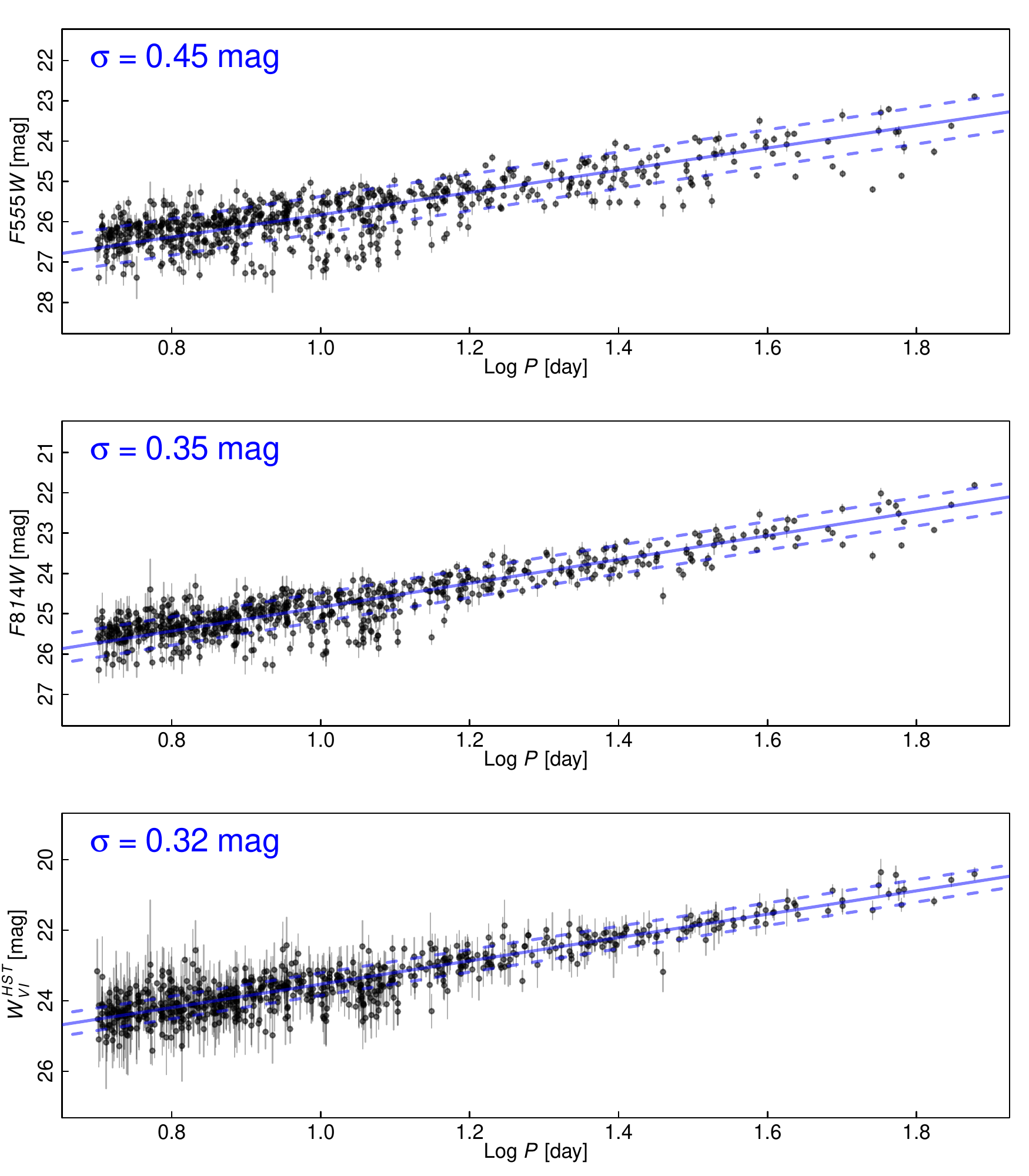}
\caption{The \ng Cepheid PLRs in \hstvs, \hstis, and \wesens. The solid blue lines indicate the best-fit PLRs, while the dashed blue lines indicate their 1$\sigma$ scatter.}\label{fig_plr}
\end{figure}

\subsection{Cepheid PLRs}

Cepheids are known to follow a linear relationship between their absolute magnitudes and the logarithm of their periods. We fit PLRs to our sample using
\[
m = a\, \log P + b,
\]
where $m$ is the Cepheid apparent magnitude (\hstv or \hstis) or Wesenheit index (\wesens), $P$ is the Cepheid period in units of days, $a$ is the PLR slope, and $b$ is the intercept. We adopted the PLR slopes derived by \citet{2019ApJ...876...85R} while setting the intercepts as free parameters. We weighted individual Cepheids by the inverse of the sum of squared measurement errors and the square of the PLR scatter determined by \citet[][]{2019ApJ...876...85R}: 0.312, 0.202, and 0.085\,mag for \hstvs, \hstis, and \wesens, respectively. For the \wesen index, we also tried two additional PLR fits: in one fit we kept both the slope and intercept as free parameters, while in the other fit we limited our sample to the same period range of the LMC sample studied by \citet{2019ApJ...876...85R}. We note that our completeness cut of $P > 5$\,d also eliminated the contamination from overtone Cepheids, whose periods are typically shorter than 5 days. The \hstvs, \hstis, and \wesen PLRs are shown in Figure~\ref{fig_plr} and their best-fit values are presented in Table~\ref{tab_plr}.

\begin{deluxetable*}{cccccccccc}
\tablecaption{Cepheid PLRs\label{tab_plr}}
\tablewidth{0pt}
\tablehead{
\colhead{Band} & \colhead{Anchor} & \colhead{$P$ range} & \colhead{Slope$^a$} & \colhead{Intercept} & \colhead{scatter} & \colhead{$N$} & \colhead{Abs. ZP$^b$} & \colhead{[O/H]$^c$} & \colhead{Ref.}}
\startdata
$W_{VI}^{HST}$ &      LMC &        6--47 &          -3.310 &    15.935$\pm$0.010 &     0.085 &    68 &    -2.542$\pm$0.028 &      8.40 &   Riess+19 \\
$\pmb{W_{VI}^{HST}}$ & {\bf    N4258} & {\bf        5--75} & {\bf          -3.310} & {\bf    26.842$\pm$0.012} & {\bf     0.319} & {\bf   669} & {\bf    -2.556$\pm$0.034} & {\bf      8.63} & {\bf  this work} \\
$W_{VI}^{HST}$ &    N4258 &        5--75 &    -3.294$\pm$0.042 &    26.823$\pm$0.049 &     0.319 &   669 &    -2.574$\pm$0.034 &      8.63 &  this work \\
$W_{VI}^{HST}$ &    N4258 &        6--47 &          -3.310 &    26.849$\pm$0.014 &     0.316 &   548 &    -2.548$\pm$0.035 &      8.63 &  this work \\
   \hstv &      LMC &        6--47 &          -2.760 &    17.638$\pm$0.038 &     0.312 &    68 &    -0.839$\pm$0.046 &      8.40 &   Riess+19 \\
   \hstv &    N4258 &        5--75 &          -2.760 &    28.586$\pm$0.018 &     0.453 &   669 &    -0.811$\pm$0.037 &      8.63 &  this work \\
   \hsti &      LMC &        6--47 &          -2.960 &    16.854$\pm$0.025 &     0.202 &    68 &    -1.623$\pm$0.036 &      8.40 &   Riess+19 \\
   \hsti &    N4258 &        5--75 &          -2.960 &    27.798$\pm$0.014 &     0.353 &   669 &    -1.600$\pm$0.035 &      8.63 &  this work
\enddata
\tablecomments{$^a$The \ng slopes are either fixed to the LMC values (without giving uncertainty) or computed as free parameters (with uncertainty given).
$^b$Absolute PLR zeropoints. Their uncertainties are computed as $({\sigma_\mu^2 + \sigma_\mathrm{PLR}^2/(N-1)})^{1/2}$, where $\sigma_\mu$ is the distance modulus uncertainty and $\sigma_\mathrm{PLR}$ is the PLR scatter. For \hstv and \hsti PLRs, dust extinction is not included in the computation.
$^c$The \ng metallicity is discussed in \S4.2 and the Appendix. The LMC metallicity is derived by \citet{2021arXiv211008860R}.}
\end{deluxetable*}

\begin{figure}
\epsscale{1.2}
\plotone{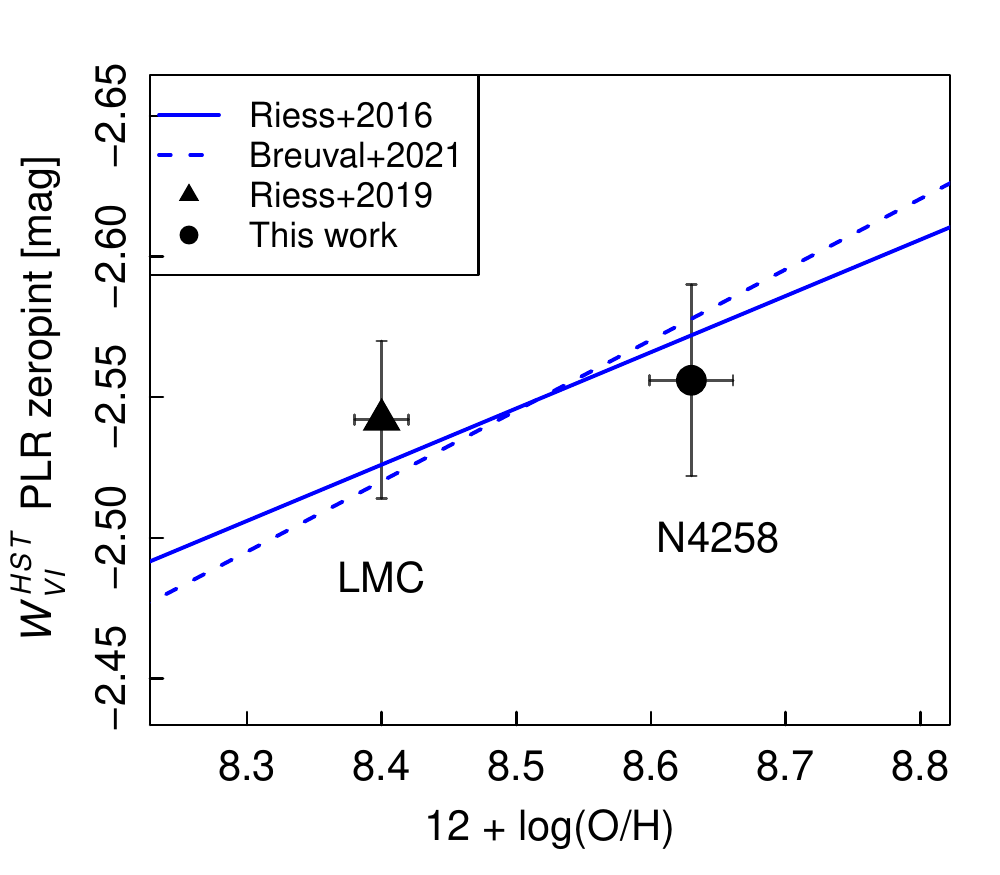}
\caption{Comparison of the absolute \wesen PLR zeropoints between our (filled circle) and \citet[][triangle]{2019ApJ...876...85R} results. The solid and dashed blue lines indicate the \wesens-band PLR metallicity dependence determined by \citet{2016ApJ...826...56R} and \citet{2021ApJ...913...38B}, respectively. We note that the \citet{2021ApJ...913...38B} result is based on ground-based $V$ and $I$ filters, whose response functions are similar but not identical to those of the \hst \hstv and \hsti filters, respectively.}\label{fig_zpz}
\end{figure}

To derive the absolute PLR zeropoints, we adopted the distance modulus of $29.397\pm0.032$\,mag to \ng  based on the recent remodeling of the water-maser motions in this system \citep{2019ApJ...886L..27R}. The resulting absolute zeropoints of the PLRs are presented in Table~\ref{tab_plr} as well. Since both the mean extinction toward the LMC and \ng may be different and the metallicity dependence of \hstv and \hsti PLRs are unavailable (most existing optical metallicity effect studies are based on non-\hst filters), we only compared our reddening-free \wesen index PLR with the corresponding value from \citet{2019ApJ...876...85R}. The \citet{2019ApJ...876...85R} calibration is based on the LMC Cepheids and detached eclipsing binary distance \citep{2019Natur.567..200P} and thus is completely independent of our result. We applied a metallicity (oxygen abundance $Z = \log$(O/H)) dependence of $-0.20\pm0.05$\,mag\,dex$^{-1}$ based on \citet{2016ApJ...826...56R} to the mean $Z$ difference between the LMC and \ng Cepheids and obtained a metallicity correction of 0.046\,mag. With this metallicity correction, we found excellent agreement ($0.032\pm0.044$\,mag) between our absolute \wesen PLR and the \citet{2019ApJ...876...85R} one, as shown in Figure~\ref{fig_zpz}. If we assume the metallicity dependence of $-0.25\pm0.06$\,mag\,dex$^{-1}$ derived using ground-based filters by \citet{2021ApJ...913...38B}, the agreement is $0.044\pm0.044$\,mag.

\subsection{Metallicity Dependence}

A robust determination of the Cepheid PLR metallicity dependence requires high precision and consistent calibrations in the chemical abundance measurement, Cepheid photometry, and dust extinction estimation simultaneously. Because our \ng Cepheids cover a relatively large range of radial distances from the galactic center, it is possible to study the metallicity effect using only the abundance gradient of the galaxy and bypassing the abundance zeropoint error among \ng and other galaxies. Taking advantage of this opportunity, here we attempt to solve for the \wesens-band PLR metallicity dependence using the \ng Cepheids alone and characterize its associated uncertainty.

We acquired 37 new metallicity measurements for 26 \ion{H}{2} regions in \ng using the Low-Resolution Imaging Spectrometer \citep[LRIS;][]{1995PASP..107..375O} on the Keck-I 10\,m telescope and the Kast double spectrograph \citep{MillerStone93} on the Shane 3\,m telescope at Lick Observatory, and combined them with those from \citet[][hereafter B11]{2011ApJ...729...56B} to obtain a total of 52 \ion{H}{2} regions. For the new measurements, we followed the observation strategy and analysis methods described in \S2.5 of \citet{2005ApJ...627..579R} to derive the metallicities of those \ion{H}{2} regions. There are 10 \ion{H}{2} regions in common between our new measurements and the B11 sample, and we averaged them with weighted means. We computed the oxygen abundances $Z = 12 + \log$(O/H) from the $R_{23}$ quantity using the \citet[][hereafter Z94]{1994ApJ...420...87Z} method and transformed them to the \citet[][hereafter PP04]{2004MNRAS.348L..59P} scale using the \citet{2021MNRAS.503.1082T} analytical solution. The PP04 scale better matches the zeropoint of the LMC Cepheid metallicity measurements \citep{2021arXiv211008860R}, which are calibrated to the solar-metallicity determinations of \citet{2009ARA&A..47..481A}. The combined \ion{H}{2} region metallicity measurements and the metallicity gradient are presented in the Appendix (Table~\ref{tab_z} and Figure~\ref{fig_zg}, respectively). 

We deprojected the galaxy for radial distances from the galactic center using the following position parameters that were adopted by B11:

\[
\left\{
     \begin{array}{ll}
       \mathrm{Distance:} & D=7.2 \,\, \mathrm{Mpc} \\
       \mathrm{Center\, R.A.\ (J2000):}  & \alpha_0=12^h18^m57.5^s \\
       \mathrm{Center\, Dec.\ (J2000):} & \delta_0=47^\circ18\arcmin 14.3\arcsec \\
       \mathrm{Position\, angle:}  & \phi=150^\circ \\
       \mathrm{Inclination:} & i=72^\circ .
     \end{array}
\right.
\]
We found a $Z$ gradient of $9.005 - 0.015\,r$ on the Z94 scale and a gradient of $8.771 - 0.014\,r$ on the PP04 scale, where $r$ is the radial distance from the galactic center in  units of kpc. The scatter of these $Z$ gradient relations is $\sim 0.14$ dex, which is likely to have an astrophysical origin rather than measurement errors. We then inferred the Cepheid metallicity using the gradient in the PP04 scale. We note that the Cepheid metallicity estimation is independent of the adopted \ng distance for deprojection.

\begin{figure}[t]
\epsscale{1.2}
\plotone{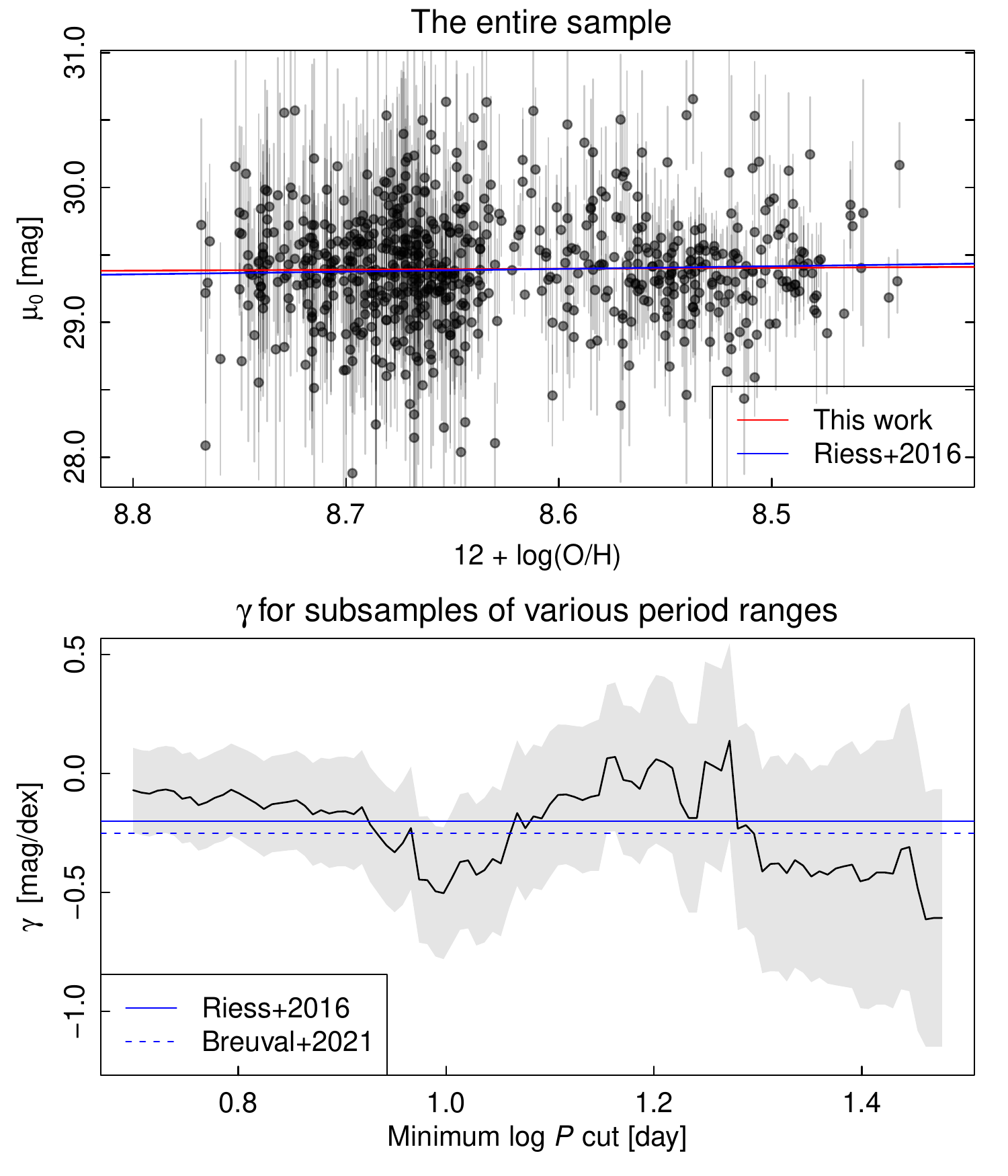}
\caption{{\it Top}: \ng distance moduli based on individual Cepheids against the Cepheid oxygen abundances estimated from their radial locations. The red line indicates the best-fit regression result using the entire sample of this work, while the blue line shows the \citet{2016ApJ...826...56R} determination. {\it Bottom}: The best-fit $\gamma$ values from samples with various minimum period cuts (black curve), with the gray shaded region indicating their random errors. The solid and dashed blue lines indicate the \citet{2016ApJ...826...56R} and \citet{2021ApJ...913...38B} determinations, respectively.}\label{fig_met}
\end{figure}

To derive the metallicity dependence $\gamma = {\mathrm{d}ZP}/{\mathrm{d}Z}$, we performed a linear regression fit between the \wesen PLR intercepts of individual Cepheids and the metallicity $Z$ calculated from the Cepheid locations, as shown in the top panel of Figure~\ref{fig_met}. We calculated $\gamma$ with a variety of minimum period cuts of the Cepheid sample, and found that our determination on this term is rather unstable owing to random errors, as shown in the bottom panel of Figure~\ref{fig_met}. Using the entire Cepheid sample, the best fit yielded $\gamma = -0.07 \pm 0.18$\,mag\,dex$^{-1}$, where the quoted uncertainty only includes a random error. The large random error in our measurement was caused by (1) the shallow \ng metallicity gradient ($\sim 4$ times shallower than the Milky Way gradient), (2) the large scatter of the abundances, and (3) the Cepheid \wesen measurement errors.

We further characterized the leading systematic error in our metallicity dependence estimate. There is a disagreement regarding the \ng metallicity gradient in the recent literature. For example, B11 obtained a gradient of $Z=8.87(\pm0.02)-0.012(\pm0.002)\,r$ using the M91 calibration and $Z=8.49(\pm0.02) - 0.011(\pm0.002)\,r$ using the P05 calibration, while \citet{2016ApJ...830...10H} derived a gradient of $Z = 9.035(\pm0.022) - 0.018(\pm0.002)\,r$ on the Z94 scale. When measuring the differential term $\gamma$, only the slope of these relations (which marginally depends on the adopted scales) matters. We took the slope difference of 0.006\,dex\,kpc$^{-1}$ among these measurements and obtained a mean systematic error in $\gamma$ of 0.11\,mag\,dex$^{-1}$ for various minimum period cuts. Combining the random and systematic errors, we found that $\gamma = -0.07 \pm 0.21$\,mag\,dex$^{-1}$. The large uncertainty in our metallicity-effect estimate prevents us from drawing any meaningful comparisons with other studies on this matter. We conclude that \ng itself is not a good target for a differential constraint of the Cepheid PLR metallicity effect.

\subsection{Photometry Comparison  with Previous Work}

We compared the Cepheid photometry between this work and that of \citet{2006ApJ...652.1133M} as follows. We recovered their \hstv and \hsti magnitudes (using their Table~4 and reversing their transformation equations for $V$ and $I$) and compared them to our equivalent measurements (aperture-corrected, CTE-corrected values). We found $3\sigma$-clipped mean differences for the equivalent mean magnitudes of $0.01\pm0.01$, $-0.01\pm0.01$, and $0.02\pm0.01$\,mag in \hstvs, \hstis, and \hstvs$-$\hstis, respectively.

One notable difference between the previous work and our analysis is our inclusion of extensive artificial-star simulations to correct for crowding for individual Cepheids, which was previously neglected in optical analyses. Taking this into account, we found $3\sigma$-clipped mean differences with the previous work of $-0.08\pm0.01$, $-0.14\pm0.01$, and $+0.07\pm0.01$\,mag in \hstvs, \hstis, and \hstvs$-$\hstis, respectively (our corrected values are fainter and bluer). Most of the Cepheids from \citet{2006ApJ...652.1133M} come from their ``inner'' field, which is among the densest and optically reddest fields analyzed by the SH0ES project. We note that for the SH0ES SN hosts (where the background stellar population is less red), the color shift from crowding corrections compared to previous work \citep{2016ApJ...830...10H} is much smaller, with a mean in \hstvs$-$\hsti of $\sim 0.01$\,mag \citep{Yuan2022b}.

\section{Summary}

As part of the effort in pushing the \h precision close to the 1\% level, we expanded the Cepheid search in \ngs, one of the critically important galaxies for the absolute calibration of the Cepheid distance scale. We obtained time-series observations of four \ng fields using \hst and carried out a Cepheid search with a combination of new and archival data, yielding a total of 669 Cepheid candidates. We calibrated the Cepheid PLRs using the latest geometric distance to \ng and found excellent agreement with the independently calibrated Cepheid PLR by \citet{2019ApJ...876...85R}. Finally, we studied the effect of metallicity on Cepheid PLRs using our \ng Cepheids. However, as the data only loosely constrain the PLR metallicity dependence, we conclude that \ng itself is not suitable for a robust differential determination of the Cepheid metallicity effect.

This work led to a factor of $\sim3$ increase in the NIR Cepheid sample in \ng that was analyzed by \citet{2021arXiv211204510R}. Although including \ng as a distance anchor does not significantly reduce the total \h uncertainty (as there exist two other anchors and the uncertainty has many sources), the \ng Cepheid data provided robust tests in the crowding corrections and metallicity dependence \citep{2021arXiv211204510R}. In the future, we plan to refine the \ng Cepheid photometry with the {\it James Webb Space Telescope} (\jwsts) by further evaluating the field crowding at NIR wavelengths, and extend the absolute Cepheid PLR calibrations to the \jwst photometric system.

\acknowledgments

We thank the anonymous referee for their constructive comments and suggestions. Brad Tucker designed
the multislit masks used for the Keck/LRIS spectroscopy of \ion{H}{2} regions in \ngs. Support for \hst program GO-16198 was provided by NASA through a grant from the Space Telescope Science Institute (STScI), which is operated by the Association of Universities for Research in Astronomy, Inc., under NASA contract NAS5-26555. A.V.F.'s group at U.C. Berkeley is also grateful for financial assistance from the TABASGO Foundation, the Christopher R. Redlich Fund, the Miller Institute for Basic Research in Science (in which he is a Miller Senior Fellow), and numerous individual donors. 

Some of the data presented herein were obtained at the W. M. Keck
Observatory, which is operated as a scientific partnership among the
California Institute of Technology, the University of California, and
NASA; the observatory was made possible by the generous financial
support of the W. M. Keck Foundation. The Shane/Kast red CCD detector upgrade at Lick, led by
B. Holden, was made
possible by the Heising-Simons Foundation, William and Marina Kast, and the
University of California Observatories. Research at Lick Observatory is partially supported by a
generous gift from Google.

Most of the data presented in this paper were obtained from the Mikulski Archive for Space Telescopes (MAST) at STScI. The specific observations analyzed can be accessed via \dataset[10.17909/e5sn-yr35]{https://doi.org/10.17909/e5sn-yr35}.

\newpage

\bibliographystyle{aasjournal}
\bibliography{ref}

\begin{thebibliography}{}
\expandafter\ifx\csname natexlab\endcsname\relax\def\natexlab#1{#1}\fi
\providecommand{\url}[1]{\href{#1}{#1}}
\providecommand{\dodoi}[1]{doi:~\href{http://doi.org/#1}{\nolinkurl{#1}}}
\providecommand{\doeprint}[1]{\href{http://ascl.net/#1}{\nolinkurl{http://ascl.net/#1}}}
\providecommand{\doarXiv}[1]{\href{https://arxiv.org/abs/#1}{\nolinkurl{https://arxiv.org/abs/#1}}}

\bibitem[{{Asplund} {et~al.}(2009){Asplund}, {Grevesse}, {Sauval}, \&
  {Scott}}]{2009ARA&A..47..481A}
{Asplund}, M., {Grevesse}, N., {Sauval}, A.~J., \& {Scott}, P. 2009, \araa, 47,
  481, \dodoi{10.1146/annurev.astro.46.060407.145222}

\bibitem[{{Bresolin}(2011)}]{2011ApJ...729...56B}
{Bresolin}, F. 2011, \apj, 729, 56, \dodoi{10.1088/0004-637X/729/1/56}

\bibitem[{{Breuval} {et~al.}(2021){Breuval}, {Kervella}, {Wielg{\'o}rski},
  {Gieren}, {Graczyk}, {Trahin}, {Pietrzy{\'n}ski}, {Arenou}, {Javanmardi}, \&
  {Zgirski}}]{2021ApJ...913...38B}
{Breuval}, L., {Kervella}, P., {Wielg{\'o}rski}, P., {et~al.} 2021, \apj, 913,
  38, \dodoi{10.3847/1538-4357/abf0ae}

\bibitem[{{Cardelli} {et~al.}(1989){Cardelli}, {Clayton}, \&
  {Mathis}}]{1989ApJ...345..245C}
{Cardelli}, J.~A., {Clayton}, G.~C., \& {Mathis}, J.~S. 1989, \apj, 345, 245,
  \dodoi{10.1086/167900}

\bibitem[{{Deustua} {et~al.}(2017){Deustua}, {Mack}, {Bajaj}, \&
  {Khandrika}}]{2017wfc..rept...14D}
{Deustua}, S.~E., {Mack}, J., {Bajaj}, V., \& {Khandrika}, H. 2017, {WFC3/UVIS
  Updated 2017 Chip-Dependent Inverse Sensitivity Values}, Space Telescope WFC
  Instrument Science Report

\bibitem[{{Di Valentino} {et~al.}(2021){Di Valentino}, {Mena}, {Pan},
  {Visinelli}, {Yang}, {Melchiorri}, {Mota}, {Riess}, \&
  {Silk}}]{2021CQGra..38o3001D}
{Di Valentino}, E., {Mena}, O., {Pan}, S., {et~al.} 2021, Classical and Quantum
  Gravity, 38, 153001, \dodoi{10.1088/1361-6382/ac086d}

\bibitem[{{Fausnaugh} {et~al.}(2015){Fausnaugh}, {Kochanek}, {Gerke}, {Macri},
  {Riess}, \& {Stanek}}]{2015MNRAS.450.3597F}
{Fausnaugh}, M.~M., {Kochanek}, C.~S., {Gerke}, J.~R., {et~al.} 2015, \mnras,
  450, 3597, \dodoi{10.1093/mnras/stv881}

\bibitem[{{Freedman} {et~al.}(2001){Freedman}, {Madore}, {Gibson}, {Ferrarese},
  {Kelson}, {Sakai}, {Mould}, {Kennicutt}, {Ford}, {Graham}, {Huchra},
  {Hughes}, {Illingworth}, {Macri}, \& {Stetson}}]{2001ApJ...553...47F}
{Freedman}, W.~L., {Madore}, B.~F., {Gibson}, B.~K., {et~al.} 2001, \apj, 553,
  47, \dodoi{10.1086/320638}

\bibitem[{{Fruchter} \& {Hook}(2002)}]{2002PASP..114..144F}
{Fruchter}, A.~S., \& {Hook}, R.~N. 2002, \pasp, 114, 144,
  \dodoi{10.1086/338393}

\bibitem[{{Gonzaga} {et~al.}(2012){Gonzaga}, {Hack}, {Fruchter}, \&
  {Mack}}]{2012drzp.book.....G}
{Gonzaga}, S., {Hack}, W., {Fruchter}, A., \& {Mack}, J. 2012, {The DrizzlePac
  Handbook}

\bibitem[{{Hoffmann} \& {Macri}(2015)}]{2015AJ....149..183H}
{Hoffmann}, S.~L., \& {Macri}, L.~M. 2015, \aj, 149, 183,
  \dodoi{10.1088/0004-6256/149/6/183}

\bibitem[{{Hoffmann} {et~al.}(2016){Hoffmann}, {Macri}, {Riess}, {Yuan},
  {Casertano}, {Foley}, {Filippenko}, {Tucker}, {Chornock}, {Silverman},
  {Welch}, {Goobar}, \& {Amanullah}}]{2016ApJ...830...10H}
{Hoffmann}, S.~L., {Macri}, L.~M., {Riess}, A.~G., {et~al.} 2016, \apj, 830,
  10, \dodoi{10.3847/0004-637X/830/1/10}

\bibitem[{{Hubble}(1929)}]{1929PNAS...15..168H}
{Hubble}, E. 1929, Proceedings of the National Academy of Science, 15, 168,
  \dodoi{10.1073/pnas.15.3.168}

\bibitem[{{Hubble}(1926)}]{1926ApJ....63..236H}
{Hubble}, E.~P. 1926, \apj, 63, 236, \dodoi{10.1086/142976}

\bibitem[{{Klagyivik} \& {Szabados}(2009)}]{2009A&A...504..959K}
{Klagyivik}, P., \& {Szabados}, L. 2009, \aap, 504, 959,
  \dodoi{10.1051/0004-6361/200811464}

\bibitem[{{Leavitt} \& {Pickering}(1912)}]{1912HarCi.173....1L}
{Leavitt}, H.~S., \& {Pickering}, E.~C. 1912, Harvard College Observatory
  Circular, 173, 1

\bibitem[{{Macri} {et~al.}(2006){Macri}, {Stanek}, {Bersier}, {Greenhill}, \&
  {Reid}}]{2006ApJ...652.1133M}
{Macri}, L.~M., {Stanek}, K.~Z., {Bersier}, D., {Greenhill}, L.~J., \& {Reid},
  M.~J. 2006, \apj, 652, 1133, \dodoi{10.1086/508530}

\bibitem[{{Miller} \& {Stone}(1993)}]{MillerStone93}
{Miller}, J.~S., \& {Stone}, R.~P.~S. 1993, {Lick Obs. Tech. Rep. 66 (Santa
  Cruz, CA: Lick Observatory)}

\bibitem[{{Newman} {et~al.}(2001){Newman}, {Ferrarese}, {Stetson}, {Maoz},
  {Zepf}, {Davis}, {Freedman}, \& {Madore}}]{2001ApJ...553..562N}
{Newman}, J.~A., {Ferrarese}, L., {Stetson}, P.~B., {et~al.} 2001, \apj, 553,
  562, \dodoi{10.1086/320969}

\bibitem[{{Oke} {et~al.}(1995){Oke}, {Cohen}, {Carr}, {Cromer}, {Dingizian},
  {Harris}, {Labrecque}, {Lucinio}, {Schaal}, {Epps}, \&
  {Miller}}]{1995PASP..107..375O}
{Oke}, J.~B., {Cohen}, J.~G., {Carr}, M., {et~al.} 1995, \pasp, 107, 375,
  \dodoi{10.1086/133562}

\bibitem[{{Pettini} \& {Pagel}(2004)}]{2004MNRAS.348L..59P}
{Pettini}, M., \& {Pagel}, B. E.~J. 2004, \mnras, 348, L59,
  \dodoi{10.1111/j.1365-2966.2004.07591.x}

\bibitem[{{Pietrzy{\'n}ski} {et~al.}(2019){Pietrzy{\'n}ski}, {Graczyk},
  {Gallenne}, {Gieren}, {Thompson}, {Pilecki}, {Karczmarek}, {G{\'o}rski},
  {Suchomska}, {Taormina}, {Zgirski}, {Wielg{\'o}rski}, {Ko{\l}aczkowski},
  {Konorski}, {Villanova}, {Nardetto}, {Kervella}, {Bresolin}, {Kudritzki},
  {Storm}, {Smolec}, \& {Narloch}}]{2019Natur.567..200P}
{Pietrzy{\'n}ski}, G., {Graczyk}, D., {Gallenne}, A., {et~al.} 2019, \nat, 567,
  200, \dodoi{10.1038/s41586-019-0999-4}

\bibitem[{{Reid} {et~al.}(2019){Reid}, {Pesce}, \&
  {Riess}}]{2019ApJ...886L..27R}
{Reid}, M.~J., {Pesce}, D.~W., \& {Riess}, A.~G. 2019, \apjl, 886, L27,
  \dodoi{10.3847/2041-8213/ab552d}

\bibitem[{{Riess} {et~al.}(2021{\natexlab{a}}){Riess}, {Casertano}, {Yuan},
  {Bowers}, {Macri}, {Zinn}, \& {Scolnic}}]{2021ApJ...908L...6R}
{Riess}, A.~G., {Casertano}, S., {Yuan}, W., {et~al.} 2021{\natexlab{a}},
  \apjl, 908, L6, \dodoi{10.3847/2041-8213/abdbaf}

\bibitem[{{Riess} {et~al.}(2019){Riess}, {Casertano}, {Yuan}, {Macri}, \&
  {Scolnic}}]{2019ApJ...876...85R}
{Riess}, A.~G., {Casertano}, S., {Yuan}, W., {Macri}, L.~M., \& {Scolnic}, D.
  2019, \apj, 876, 85, \dodoi{10.3847/1538-4357/ab1422}

\bibitem[{{Riess} {et~al.}(2005){Riess}, {Li}, {Stetson}, {Filippenko}, {Jha},
  {Kirshner}, {Challis}, {Garnavich}, \& {Chornock}}]{2005ApJ...627..579R}
{Riess}, A.~G., {Li}, W., {Stetson}, P.~B., {et~al.} 2005, \apj, 627, 579,
  \dodoi{10.1086/430497}

\bibitem[{{Riess} {et~al.}(2009){Riess}, {Macri}, {Casertano}, {Sosey},
  {Lampeitl}, {Ferguson}, {Filippenko}, {Jha}, {Li}, {Chornock}, \&
  {Sarkar}}]{2009ApJ...699..539R}
{Riess}, A.~G., {Macri}, L., {Casertano}, S., {et~al.} 2009, \apj, 699, 539,
  \dodoi{10.1088/0004-637X/699/1/539}

\bibitem[{{Riess} {et~al.}(2011){Riess}, {Macri}, {Casertano}, {Lampeitl},
  {Ferguson}, {Filippenko}, {Jha}, {Li}, \& {Chornock}}]{2011ApJ...730..119R}
---. 2011, \apj, 730, 119, \dodoi{10.1088/0004-637X/730/2/119}

\bibitem[{{Riess} {et~al.}(2016){Riess}, {Macri}, {Hoffmann}, {Scolnic},
  {Casertano}, {Filippenko}, {Tucker}, {Reid}, {Jones}, {Silverman},
  {Chornock}, {Challis}, {Yuan}, {Brown}, \& {Foley}}]{2016ApJ...826...56R}
{Riess}, A.~G., {Macri}, L.~M., {Hoffmann}, S.~L., {et~al.} 2016, \apj, 826,
  56, \dodoi{10.3847/0004-637X/826/1/56}

\bibitem[{{Riess} {et~al.}(2021{\natexlab{b}}){Riess}, {Yuan}, {Macri},
  {Scolnic}, {Brout}, {Casertano}, {Jones}, {Murakami}, {Breuval}, {Brink},
  {Filippenko}, {Hoffmann}, {Jha}, {Kenworthy}, {Mackenty}, {Stahl}, \&
  {Zheng}}]{2021arXiv211204510R}
{Riess}, A.~G., {Yuan}, W., {Macri}, L.~M., {et~al.} 2021{\natexlab{b}}, arXiv
  e-prints, arXiv:2112.04510.
\newblock \doarXiv{2112.04510}

\bibitem[{{Ripepi} {et~al.}(2021){Ripepi}, {Catanzaro}, {Molinaro}, {Gatto},
  {De Somma}, {Marconi}, {Romaniello}, {Leccia}, {Musella}, {Trentin},
  {Clementini}, {Testa}, {Cusano}, \& {Storm}}]{2021MNRAS.508.4047R}
{Ripepi}, V., {Catanzaro}, G., {Molinaro}, R., {et~al.} 2021, \mnras, 508,
  4047, \dodoi{10.1093/mnras/stab2460}

\bibitem[{{Ripepi} {et~al.}(2022){Ripepi}, {Catanzaro}, {Clementini}, {De
  Somma}, {Drimmel}, {Leccia}, {Marconi}, {Molinaro}, {Musella}, \&
  {Poggio}}]{2022arXiv220101126R}
{Ripepi}, V., {Catanzaro}, G., {Clementini}, G., {et~al.} 2022, arXiv e-prints,
  arXiv:2201.01126.
\newblock \doarXiv{2201.01126}

\bibitem[{{Romaniello} {et~al.}(2021){Romaniello}, {Riess}, {Mancino},
  {Anderson}, {Freudling}, {Kudritzki}, {Macri}, {Mucciarelli}, \&
  {Yuan}}]{2021arXiv211008860R}
{Romaniello}, M., {Riess}, A., {Mancino}, S., {et~al.} 2021, arXiv e-prints,
  arXiv:2110.08860.
\newblock \doarXiv{2110.08860}

\bibitem[{{Scolnic} {et~al.}(2018){Scolnic}, {Jones}, {Rest}, {Pan},
  {Chornock}, {Foley}, {Huber}, {Kessler}, {Narayan}, {Riess}, {Rodney},
  {Berger}, {Brout}, {Challis}, {Drout}, {Finkbeiner}, {Lunnan}, {Kirshner},
  {Sanders}, {Schlafly}, {Smartt}, {Stubbs}, {Tonry}, {Wood-Vasey}, {Foley},
  {Hand}, {Johnson}, {Burgett}, {Chambers}, {Draper}, {Hodapp}, {Kaiser},
  {Kudritzki}, {Magnier}, {Metcalfe}, {Bresolin}, {Gall}, {Kotak}, {McCrum}, \&
  {Smith}}]{2018ApJ...859..101S}
{Scolnic}, D.~M., {Jones}, D.~O., {Rest}, A., {et~al.} 2018, \apj, 859, 101,
  \dodoi{10.3847/1538-4357/aab9bb}

\bibitem[{{Stetson}(1987)}]{1987PASP...99..191S}
{Stetson}, P.~B. 1987, \pasp, 99, 191, \dodoi{10.1086/131977}

\bibitem[{{Stetson}(1994)}]{1994PASP..106..250S}
---. 1994, \pasp, 106, 250, \dodoi{10.1086/133378}

\bibitem[{{Stetson}(1996)}]{1996PASP..108..851S}
---. 1996, \pasp, 108, 851, \dodoi{10.1086/133808}

\bibitem[{{Teimoorinia} {et~al.}(2021){Teimoorinia}, {Jalilkhany}, {Scudder},
  {Jensen}, \& {Ellison}}]{2021MNRAS.503.1082T}
{Teimoorinia}, H., {Jalilkhany}, M., {Scudder}, J.~M., {Jensen}, J., \&
  {Ellison}, S.~L. 2021, \mnras, 503, 1082, \dodoi{10.1093/mnras/stab466}

\bibitem[{{Yoachim} {et~al.}(2009){Yoachim}, {McCommas}, {Dalcanton}, \&
  {Williams}}]{2009AJ....137.4697Y}
{Yoachim}, P., {McCommas}, L.~P., {Dalcanton}, J.~J., \& {Williams}, B.~F.
  2009, \aj, 137, 4697, \dodoi{10.1088/0004-6256/137/6/4697}

\bibitem[{{Yuan} {et~al.}(in preparation){Yuan}, {Riess}, {Brink}, {Casertano},
  {Filippenko}, {Hoffmann}, {Huang}, \& {Scolnic}}]{Yuan2022b}
{Yuan}, W.~{Macri}, L., {Riess}, A., {Brink}, T., {et~al.} in preparation

\bibitem[{{Yuan} {et~al.}(2020){Yuan}, {Fausnaugh}, {Hoffmann}, {Macri},
  {Peterson}, {Riess}, {Bentz}, {Brown}, {Bont{\`a}}, {Davies}, {Rosa},
  {Ferrarese}, {Grier}, {Hicks}, {Onken}, {Pogge}, {Storchi-Bergmann}, \&
  {Vestergaard}}]{2020ApJ...902...26Y}
{Yuan}, W., {Fausnaugh}, M.~M., {Hoffmann}, S.~L., {et~al.} 2020, \apj, 902,
  26, \dodoi{10.3847/1538-4357/abb377}

\bibitem[{{Zaritsky} {et~al.}(1994){Zaritsky}, {Kennicutt}, \&
  {Huchra}}]{1994ApJ...420...87Z}
{Zaritsky}, D., {Kennicutt}, Robert~C., J., \& {Huchra}, J.~P. 1994, \apj, 420,
  87, \dodoi{10.1086/173544}

\end{thebibliography}

\appendix

\begin{deluxetable}{lccrccccr}[H]
\tablecaption{\ion{H}{2} Region Metallicity Measurements\label{tab_z}}
\tablewidth{0pt}
\tablehead{
\colhead{ID} & \colhead{RA (J2000)} & \colhead{Dec (J2000)} & \colhead{$r$ (kpc)} & \colhead{$R_{23}$} & \colhead{$Z$ (Z94)} & \colhead{$Z$ (PP04)} & \colhead{$\sigma$} & \colhead{Source}
}
\startdata
b15 &    184.73279 &     47.30519 &      1.44 &      2.95 &      9.03 &      8.80 &      0.05 &             B11 \\
b16 &    184.73413 &     47.30311 &      1.49 &      4.37 &      8.86 &      8.63 &      0.05 &             B11 \\
b17 &    184.73450 &     47.29611 &      2.88 &      3.89 &      8.92 &      8.69 &      0.05 &             B11 \\
n12 &    184.74600 &     47.31554 &      4.01 &      3.36 &      8.98 &      8.75 &      0.06 &             new \\
b14 &    184.74383 &     47.31806 &      4.11 &      3.74 &      8.93 &      8.71 &      0.04 &        B11, new \\
n14 &    184.71637 &     47.31225 &      4.30 &      4.15 &      8.89 &      8.66 &      0.08 &             new \\
b13 &    184.70629 &     47.32233 &      5.42 &      3.83 &      8.93 &      8.70 &      0.04 &        B11, new \\
b12 &    184.74554 &     47.32517 &      6.09 &      4.68 &      8.82 &      8.60 &      0.05 &             B11 \\
 b6 &    184.71271 &     47.34814 &      6.48 &      3.31 &      8.98 &      8.76 &      0.05 &             B11 \\
 n3 &    184.72017 &     47.34598 &      6.67 &      3.89 &      8.92 &      8.69 &      0.08 &             new \\
 b5 &    184.70579 &     47.35228 &      6.92 &      3.61 &      8.95 &      8.72 &      0.04 &        B11, new \\
b11 &    184.73904 &     47.33453 &      6.94 &      3.13 &      9.01 &      8.78 &      0.03 &        B11, new \\
b10 &    184.73562 &     47.33756 &      7.02 &      3.24 &      8.99 &      8.77 &      0.05 &             B11 \\
 b9 &    184.73308 &     47.34042 &      7.23 &      3.53 &      8.96 &      8.73 &      0.04 &        B11, new \\
 b8 &    184.73000 &     47.34319 &      7.36 &      3.51 &      8.96 &      8.73 &      0.04 &        B11, new \\
 b4 &    184.70217 &     47.35689 &      7.58 &      4.36 &      8.86 &      8.63 &      0.04 &        B11, new \\
 b7 &    184.72554 &     47.34731 &      7.62 &      2.75 &      9.05 &      8.82 &      0.05 &             B11 \\
b19 &    184.76483 &     47.25294 &      7.92 &      4.37 &      8.86 &      8.63 &      0.05 &             B11 \\
b20 &    184.77158 &     47.25022 &      7.92 &      6.31 &      8.63 &      8.41 &      0.05 &             B11 \\
 b3 &    184.73204 &     47.36508 &     12.71 &      5.46 &      8.73 &      8.51 &      0.04 &        B11, new \\
n36 &    184.72037 &     47.37533 &     13.12 &      3.75 &      8.93 &      8.71 &      0.08 &             new \\
 b2 &    184.73025 &     47.36931 &     13.37 &      5.01 &      8.78 &      8.56 &      0.05 &             B11 \\
 n2 &    184.66781 &     47.32883 &     13.38 &      2.96 &      9.03 &      8.80 &      0.06 &             new \\
 b1 &    184.72213 &     47.37653 &     13.67 &      5.25 &      8.76 &      8.53 &      0.05 &             B11 \\
b23 &    184.76254 &     47.22711 &     13.79 &      6.17 &      8.65 &      8.43 &      0.05 &             B11 \\
n37 &    184.72482 &     47.37585 &     13.95 &      4.01 &      8.90 &      8.68 &      0.08 &             new \\
b21 &    184.74704 &     47.23744 &     13.97 &      4.26 &      8.87 &      8.64 &      0.05 &             B11 \\
n33 &    184.79207 &     47.20681 &     14.70 &      5.12 &      8.77 &      8.54 &      0.08 &             new \\
b27 &    184.78179 &     47.21117 &     14.79 &      4.37 &      8.86 &      8.63 &      0.05 &             B11 \\
b29 &    184.80558 &     47.20044 &     15.03 &      5.62 &      8.71 &      8.49 &      0.05 &             B11 \\
b28 &    184.79075 &     47.20444 &     15.28 &      5.13 &      8.77 &      8.54 &      0.05 &             B11 \\
n10 &    184.79571 &     47.20149 &     15.44 &      5.00 &      8.79 &      8.56 &      0.08 &             new \\
n32 &    184.79134 &     47.20305 &     15.50 &      3.57 &      8.95 &      8.73 &      0.08 &             new \\
 n9 &    184.81600 &     47.19271 &     15.97 &      8.23 &      8.40 &      8.24 &      0.08 &             new \\
b31 &    184.80283 &     47.19433 &     16.29 &      6.17 &      8.65 &      8.43 &      0.05 &             B11 \\
b18 &    184.68275 &     47.29047 &     16.35 &      3.98 &      8.91 &      8.68 &      0.05 &             B11 \\
b32 &    184.82475 &     47.18944 &     16.36 &      3.47 &      8.97 &      8.74 &      0.05 &             B11 \\
b25 &    184.84950 &     47.21756 &     16.57 &      5.13 &      8.77 &      8.54 &      0.05 &             B11 \\
b24 &    184.74621 &     47.22606 &     16.73 &      8.31 &      8.39 &      8.23 &      0.05 &             B11 \\
b26 &    184.85362 &     47.21656 &     17.24 &      5.01 &      8.78 &      8.56 &      0.05 &             B11 \\
b33 &    184.83825 &     47.18361 &     17.33 &      7.08 &      8.54 &      8.34 &      0.05 &             B11 \\
n11 &    184.85562 &     47.21184 &     17.49 &      4.56 &      8.84 &      8.61 &      0.08 &             new \\
b34 &    184.83187 &     47.18072 &     17.61 &      5.99 &      8.67 &      8.45 &      0.03 &        B11, new \\
b22 &    184.73733 &     47.22869 &     17.79 &      3.71 &      8.94 &      8.71 &      0.05 &             B11 \\
n29 &    184.83392 &     47.17906 &     17.85 &      4.83 &      8.81 &      8.58 &      0.08 &             new \\
 n1 &    184.64927 &     47.32124 &     18.94 &      5.44 &      8.73 &      8.51 &      0.06 &             new \\
b35 &    184.79721 &     47.17956 &     19.71 &      5.37 &      8.74 &      8.52 &      0.05 &             B11 \\
b36 &    184.83121 &     47.16556 &     19.97 &      8.51 &      8.37 &      8.22 &      0.05 &             B11 \\
n26 &    184.63013 &     47.33662 &     21.15 &      4.14 &      8.89 &      8.66 &      0.08 &             new \\
b30 &    184.75992 &     47.19464 &     21.55 &      8.14 &      8.41 &      8.24 &      0.04 &        B11, new \\
 n7 &    184.61146 &     47.35218 &     23.36 &      3.72 &      8.94 &      8.71 &      0.06 &             new \\
 n5 &    184.62029 &     47.32604 &     25.13 &      5.59 &      8.71 &      8.49 &      0.08 &             new
\enddata
\end{deluxetable}

\begin{figure}
\epsscale{1.2}
\plotone{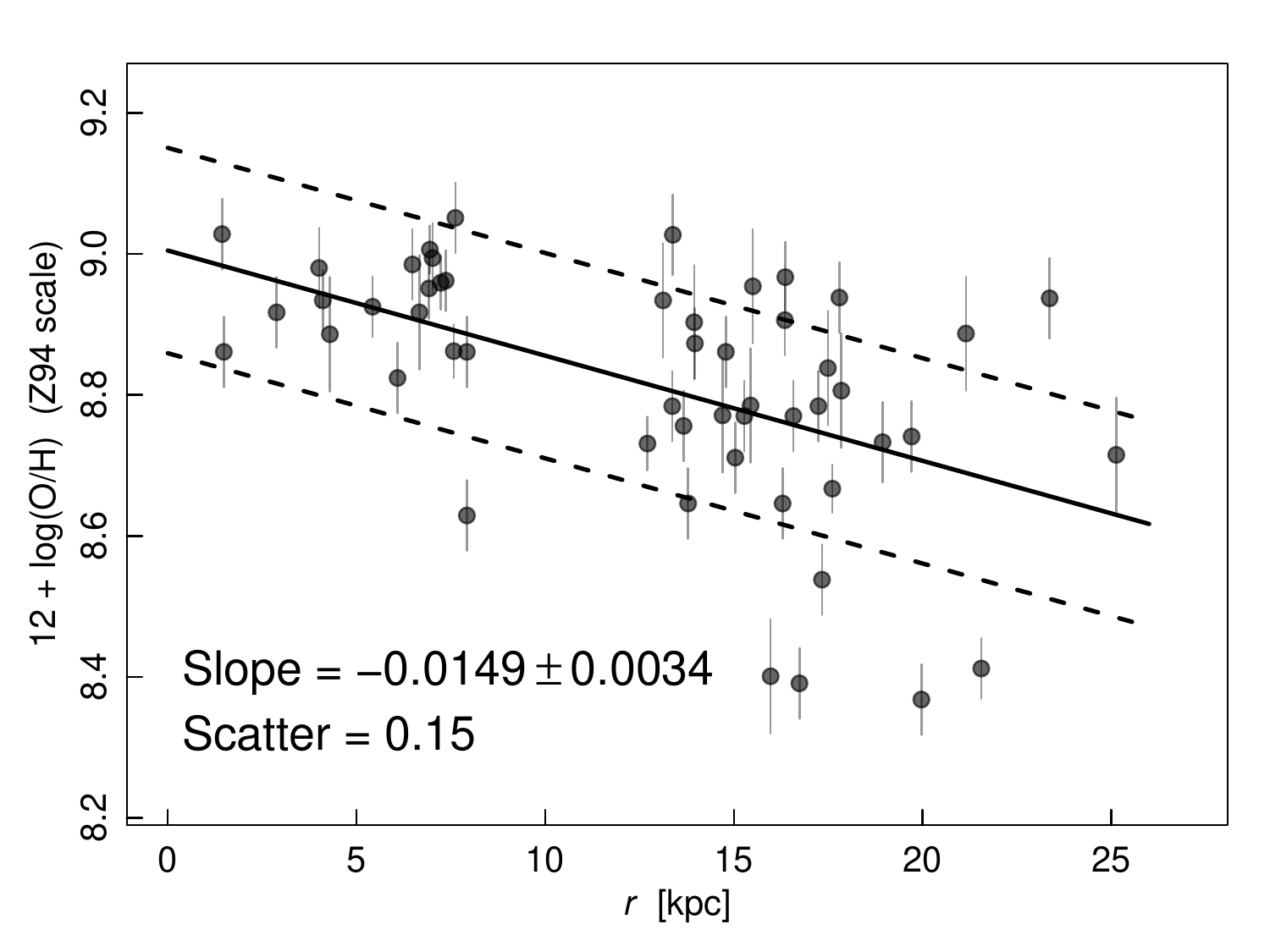}
\caption{\ion{H}{2} metallicity measurements against their deprojected radial distances from the galactic center. The solid line shows the best-fit linear relation, while the dashed lines indicate the $\pm 1 \sigma$ scatter.}\label{fig_zg}
\end{figure}

\end{document}